\documentclass[numsec,webpdf,modern,large]{oup-authoring-template}

\graphicspath{{Fig/}}

\theoremstyle{thmstyleone}%
\theoremstyle{thmstyletwo}%
\theoremstyle{thmstylethree}%

\usepackage{booktabs}
\usepackage{graphicx}
\usepackage{subfig}
\usepackage{multirow} 
\usepackage{xcolor}
\usepackage{xurl}

\begin{document}

\journaltitle{Bioinformatics}
\DOI{xxxx}
\copyrightyear{YEAR}
\pubyear{YEAR}
\vol{XX}
\issue{x}
\access{Published: Date added during production}
\appnotes{Paper}

\firstpage{1}


\title[scGPL]{Bridging Graph and Pairwise Learning for Gene Regulatory Network Inference from scRNA-seq Data}

\author[1,$\dagger$]{Hang Li}
\author[2,$\dagger$]{Jiaxin Qi}
\author[1]{Yan Cui}
\author[1]{Yuhua Zheng}
\author[1,2,$\ast$]{Jianqiang Huang\ORCID{0000-0001-5735-2910}}

\address[1]{\orgdiv{Hangzhou Institute for Advanced Study}, \orgname{University of Chinese Academy of Sciences}, \orgaddress{\street{Hangzhou}, \postcode{310024‌}, \state{Zhejiang}, \country{China}}}
\address[2]{\orgdiv{Computer Network Information Center}, \orgname{Chinese Academy of Sciences}, \orgaddress{\street{Haidian}, \postcode{100083}, \state{Beijing}, \country{China}}}

\corresp[$\dagger$]{These authors contributed equally to this work.}
\corresp[$\ast$]{Corresponding author. \href{email:jqhuang@cnic.cn}{jqhuang@cnic.cn}}

\received{Date}{0}{Year}
\revised{Date}{0}{Year}
\accepted{Date}{0}{Year}



\abstract{
\textbf{Motivation:} {\mdseries Gene regulatory networks (GRNs) capture the underlying interactions through which transcription factors (TFs) regulate genes.
Based on gene expression data, existing GRN inference approaches generally fall into two categories: graph-based methods, which model the GRN as a whole graph, and pairwise-based methods, which decompose the GRN into individual TF--target gene pairs for modeling.
However, each approach exhibits limitations that are precisely the strengths of its counterpart.
Graph-based methods tend to overfit due to their reliance on a single training graph, whereas pairwise-based methods benefit from abundant TF--target training pairs.
In contrast, pairwise-based methods overlook the global topological structure, which is essential for comprehensively capturing regulatory mechanisms.
}\\
\textbf{Results:} {\mdseries To address these limitations, we propose \textit{scGPL}, an end-to-end unified framework that jointly models global regulatory topology and local interactions.
\textit{scGPL} extracts macro-level features from the global regulatory graph and introduces a dynamic, importance-guided subgraph sampling mechanism to capture fine-grained local dependencies.
By fusing global representations with subgraph-enhanced local features, our framework bridges the gap between the two paradigms, incorporating essential whole-graph topological dependencies while retaining the data abundance advantage of pairwise methods.
Extensive experiments on seven benchmark datasets demonstrate that \textit{scGPL} consistently achieves state-of-the-art performance, with up to 13.7\% improvement over the strongest baseline, validating the effectiveness of our integrative design.}\\
\textbf{Availability and implementation:} {\mdseries The source code is maintained on GitHub (\url{https://github.com/talent-hang/scGPL}), and the datasets and versioned releases are archived at \href{https://doi.org/10.5281/zenodo.21608485}{https://doi.org/10.5281/zenodo.21608485 for reproducibility.}}
}






\maketitle

\section{Introduction}

\begin{figure}[t]
    \centering
    \includegraphics[width=\linewidth]{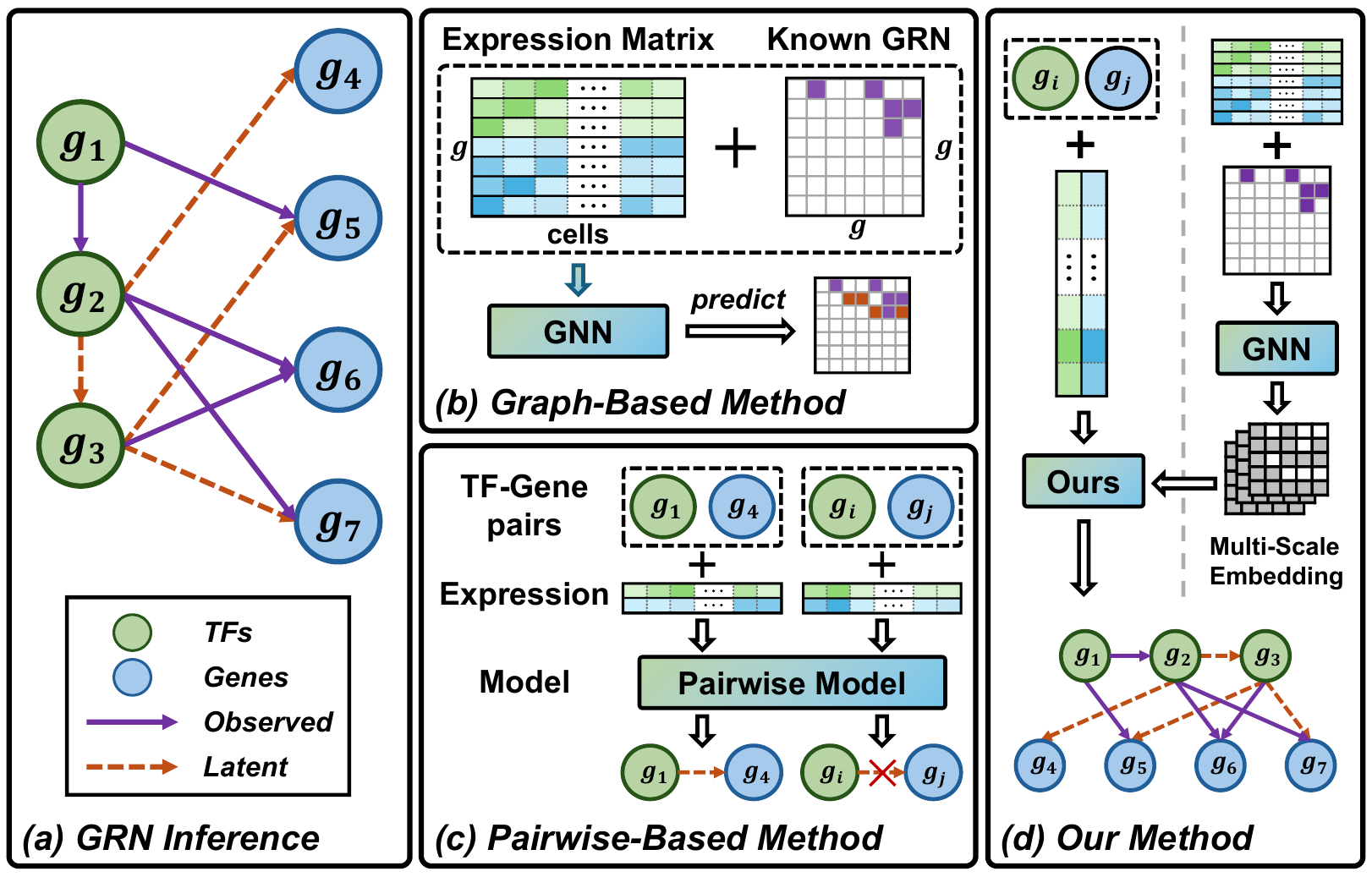}
    \caption{Illustrations of GRN inference and related approaches. (a) GRN inference aims to leverage observed regulatory relationships (purple arrow) to predict unknown ones (orange dashed arrow). (b) Graph-based methods model the entire GRN as a graph, using observed edges (purple blocks) to predict unknown edges (orange blocks) in the adjacency matrix. (c) Pairwise-based methods decompose the GRN into individual TF--gene pairs and predict their relationships independently. (d) Our method integrates both paradigms to enable comprehensive GRN inference.}
    \label{fig:figure1}
\end{figure}

Gene regulatory networks (GRNs) describe the intricate relationships by which transcription factors (TFs) control the expression of their target genes~\cite{marbach2010revealing}. A GRN is represented as a directed graph, where nodes denote TFs or genes and edges indicate regulatory relationships. As shown in Figure~\ref{fig:figure1}(a), for the transcription factor $g_1$ and gene $g_5$, the directed edge $g_1 \rightarrow g_5$ suggests that higher expression of $g_1$ leads to increased or decreased expression of $g_5$, reflecting the regulatory mechanism between them~\cite{hou2020systematic}. Thanks to the development of single-cell RNA sequencing (scRNA-seq), which provides high-throughput expression profiles of genes and TFs across individual cells, researchers can now investigate GRNs at single-cell resolution using single-cell expression data~\cite{xu2023stgrns}. 
However, current scRNA-seq datasets pose two major challenges for supervised GRN inference.
First, since obtaining reliable regulatory annotations relies on resource-intensive experimental validation, existing datasets often suffer from severely incomplete edge annotations~\cite{badia2023gene}. Second, scRNA-seq profiles tens of thousands of genes, creating a high-dimensional search space where the millions of potential regulatory interactions vastly outnumber the available annotated training samples~\cite{sun2025single}. As a result, recent studies have increasingly adopted deep learning methods for GRN inference. These methods can be broadly categorized into two representative groups: graph-based methods and pairwise-based methods.

Graph-based methods, as shown in Figure~\ref{fig:figure1}(b), treat the GRN as a whole graph, leveraging graph neural networks (GNNs) to capture global topological relationships among TFs and genes.
Specifically, GENELink~\cite{genelink} and GNNLink~\cite{gnnlink} utilize graph attention and convolutional networks to learn gene embeddings from expression data and prior topology.
Their principal strength lies in leveraging neighborhood information via message passing, which can help alleviate the impact of incomplete annotations.
For example, let $g_1$ and $g_2$ denote two different TFs, and let $g_3$ denote a target gene. If $g_1 \rightarrow g_3$ is a known regulation, and $g_2$ shares similar global topological features with $g_1$, graph-based methods can potentially learn the regulation $g_2 \rightarrow g_3$ even without explicit annotation. 
However, GNNs may obscure fine-grained local interactions due to over-smoothing during multi-hop information propagation~\cite{gronbech2020scvae}. 
To alleviate this issue, GRLGRN~\cite{wang2025grlgrn} utilizes Graph Transformers to learn implicit regulatory connections from meta-path representations of prior edges, subsequently capturing higher-order gene embeddings. 
Despite such structural enhancements, these graph-based methods, typically trained on a single GRN graph, remain prone to overfitting~\cite{gcn}. 
GRNFormer~\cite{Hegde2026GRNFormer} constructs a global co-expression network and extracts local subgraphs centered around individual TFs to perform link prediction within these subgraphs.
While this approach helps mitigate overfitting to a single graph by decomposing it into multiple TF-centric subgraphs, its extraction strategy may overlook the pair-specific influence of intermediate regulators and fail to completely capture global graph dependencies.

Pairwise-based methods, as shown in Figure~\ref{fig:figure1}(c), decompose the entire GRN into individual TF--target gene pairs and learn each regulatory interaction independently.
For instance, GNE~\cite{kc2019gne} employs multi-layer perceptrons (MLPs) to embed gene pairs, and scGREAT~\cite{wang2024scgreat} introduces a context-aware Transformer framework. 
By explicitly constructing a significantly larger training set of independent pairs, these methods successfully mitigate the overfitting problems inherent to graph-based methods and effectively capture fine-grained regulatory interactions~\cite{greener2022guide}. 
However, by isolating gene pairs, pairwise-based methods ignore the global network structure.
Consequently, they remain highly vulnerable to false negatives caused by incomplete annotations and inherently fail to capture higher-order dependencies~\cite{wang2021scgnn}.
To illustrate, if $g_1 \rightarrow g_2$ and $g_2 \rightarrow g_3$, the indirect cascade influence of $g_1$ on $g_3$ is not explicitly modeled by pairwise models. 
Ultimately, this isolated modeling paradigm hinders comprehensive network reconstruction and limits broader model applicability.


To leverage the strengths and mitigate the weaknesses of both paradigms, as shown in Figure~\ref{fig:figure1}(d), we propose \textit{scGPL} (Bridging Graph and Pairwise Learning), an end-to-end unified framework that enhances pairwise modeling with subgraph-derived topological context. \textit{scGPL} employs a dual-branch architecture that synergizes global and local views.
In the global branch, a graph neural network processes the entire gene expression matrix and prior network to extract macro-level gene embeddings, capturing hierarchical topological features.
In the local branch, we introduce a dynamic subgraph sampling mechanism to explicitly model the neighborhood context of specific interactions.
Specifically, for each candidate TF--target pair, we first initialize a local subgraph and identify a candidate frontier comprising their first-order neighbors. Through an iterative process, we calculate importance scores for frontier nodes and select the top-$k$ most informative nodes to progressively expand the subgraph structure. This strategy ensures that critical regulatory dependencies---such as co-factors or intermediate signaling genes---are included in the local representation. Finally, \textit{scGPL} fuses these subgraph-enhanced local features with the global embeddings, producing unified representations that reflect both the broad regulatory topology and the fine-grained interaction context.
This approach effectively bridges the gap between the two paradigms: it incorporates essential topological dependencies while retaining the data-abundance advantage of pairwise methods.

Our main contributions are summarized as follows:
\begin{list}{\labelitemi}{%
  \setlength{\leftmargin}{1.5em}%
  \setlength{\labelsep}{0.6em}%
  \setlength{\labelwidth}{\dimexpr\leftmargin-\labelsep\relax}%
  \setlength{\itemindent}{0pt}%
  \setlength{\listparindent}{0pt}%
  \setlength{\topsep}{0pt}%
  \setlength{\partopsep}{0pt}%
  \setlength{\itemsep}{0pt}%
  \setlength{\parsep}{0pt}%
}
  \item We systematically analyze the key challenges in GRN inference from scRNA-seq data and categorize existing methods into graph-based and pairwise-based paradigms, revealing their complementary strengths and limitations.
  \item Based on these insights, we propose \emph{scGPL}, a novel unified framework that integrates global topology modeling with subgraph-enhanced local interaction learning. This design allows the model to effectively capture both local topological structures and global network information while retaining the advantage of large-scale training samples.
  \item Extensive experiments on seven widely used GRN benchmark datasets demonstrate that \emph{scGPL} consistently outperforms state-of-the-art methods, validating the effectiveness, robustness, and scalability of our integrative design.
\end{list}

\section{Method}
\subsection{Problem Formulation}

\noindent\textbf{GRN inference}. We represent the gene regulatory network as a directed graph $(G, E)$, where $G \!=\! \{g_i\}_{i=1}^n$ denotes the set of transcription factors and target genes (i.e., the nodes), and $E \!=\! \{(g_i, g_j)\}\!\subseteq\!G\!\times\!G$ denotes the set of regulatory interactions (i.e., the edges), where $g_i \!\rightarrow \!g_j$ indicates that $g_i$ regulates $g_j$.
In practice, regulatory interactions are typically directed from TFs to their target genes; however, for notational convenience, we use $g$ to denote both TFs and genes when no ambiguity arises~\cite{gnnlink}.
The goal of GRN inference is to recover unknown edges from observational data.

Given single-cell RNA sequencing data from $c$ cells, each cell contains expression values corresponding to genes in $G$. For cell $i$, its gene expression profile is represented as $v_i \!=\! (v_{i,1}, v_{i,2}, \dots, v_{i,n})$, where $n \!=\! |G|$. The scRNA-seq data can thus be represented as the expression matrix $X \!= \! [\,v_{i,j}\,]_{i=1,\dots,c}^{j=1,\dots,n}\in \mathbb{R}^{c\times n}$.
For modeling purposes, define $y_{ij} \in \{0,1\}$ as the ground‑truth label for the edge \((g_i, g_j)\) to indicate whether $g_i$ regulates $g_j$, and split $E$ into training set $E_{\mathrm{tr}}$ and test set $E_{\mathrm{te}}$. The objective for GRN inference is to minimize:
\begin{equation}
  \mathcal{L}_{\mathrm{GRN}} = \sum_{(g_i,g_j)\,\in\,E_{\mathrm{tr}}}\mathcal{L}\left(m\left(X; E_{\mathrm{tr}}\right), y_{ij}\right),
\end{equation}
where $m$ is the trainable model designed to extract features and predict $y_{ij}$, and $\mathcal{L}$ denotes the corresponding loss function.

\begin{figure*}[t]
    \centering
    \includegraphics[width=0.87\textwidth]{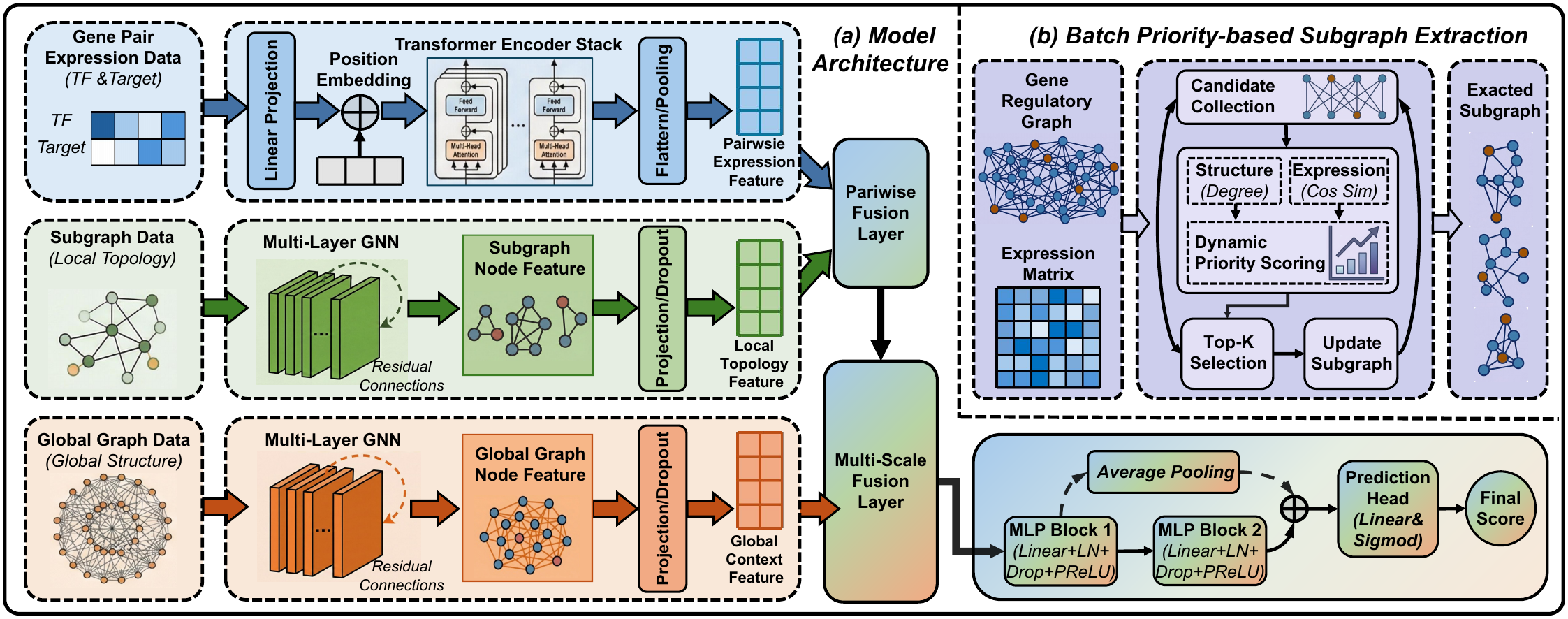}
    \caption{The overall framework of scGPL. (a) The model architecture. scGPL employs a dual-stream learning framework comprising a Pairwise Branch and a Global Branch for multi-view feature extraction. The Pairwise Branch integrates two parallel components: the Transformer Module (top), capturing co-expression dependencies from pairwise expression profiles, and the Subgraph Module (middle), extracting local topological features from dynamically sampled subgraphs. The Global Branch (bottom) uses a Global GNN to capture macro-level network structure context. Embeddings from both branches are hierarchically aggregated in the Multi-Scale Fusion Layer and processed by an MLP-based prediction head to infer gene regulatory links. (b) The Dynamic Subgraph Extraction module. This module constructs informative local subgraphs via a priority-based iterative expansion strategy. Starting from the initial gene pair, the module iteratively expands the subgraph by selecting top-$k$ neighbors based on a hybrid importance score, providing inputs to the Subgraph Module.}
    \label{fig:modelframework}
\end{figure*}

\noindent\textbf{Graph-based methods} regard the entire GRN as a single graph to train a GNN. Specifically, after initializing node embeddings directly from the expression matrix (e.g., $e_k \!=\! X_{:,k} \!\in\! \mathbb{R}^c$ for gene $g_k$), the GNN propagates these features along the training edges $E_{\mathrm{tr}}$ via message passing to capture topological dependencies. However, this simultaneous computation restricts the model to a single macroscopic training graph, leading to severe overfitting.

\noindent\textbf{Pairwise-based methods} decompose the GRN into independent TF-target pairs $(g_i,g_j)$, predicting the label $y_{ij}$ by learning a joint representation of their expression embeddings $e_i\!=\!X_{:,i}$ and $e_j\!=\!X_{:,j}$. While generating abundant training samples effectively mitigates overfitting, this formulation entirely ignores the prior network structure $E_\mathrm{tr}$, failing to model the global information essential for complex regulatory pathways.

\subsection{The scGPL Framework}

\subsubsection{Overview}
To overcome the limitations of graph-based and pairwise-based methods, we propose \textit{scGPL}, an end-to-end unified framework that seamlessly blends global regulatory topology with fine‑grained local interactions.
As illustrated in Figure~\ref{fig:modelframework}, \textit{scGPL} consists of three integrated components: 
(1) Global Branch, which extracts multi-scale topological embeddings from the entire regulatory graph to capture broad regulatory pathways;
(2) Pairwise Branch, which serves as the core of our localized modeling and comprises two sub-modules: a Transformer Module for capturing non-linear co-expression dependencies, and a dynamic Subgraph Module for encoding local regulatory contexts;
(3) Multi-scale Fusion Module, which hierarchically integrates these global and local biological perspectives to yield comprehensive and robust link predictions.

\subsubsection{Global Branch}
Given the training gene regulatory graph $(G, E_{tr})$ with $|G|\!=\!n$, the corresponding adjacency matrix $A_{tr}\!\in\!\{0,1\}^{n\times n}$, and the expression matrix $X\!\in\!\mathbb{R}^{c\times n}$, we employ a multi-layer GNN to extract hierarchical graph embeddings.
The message-passing process at the $l$-th layer ($l \ge 1$) is formulated as:
\begin{equation}
{H^{(l)} = m_g^{(l)}\left(H^{(l-1)}, A_{tr}\right),}
\end{equation}
where $H^{(0)}$ represents the initial node features derived from the expression matrix $X$, and $H^{(l)}\!\in\!\mathbb{R}^{n \times d}$ denotes the node embedding matrix at the $l$-th layer. Specifically, $d$ is the hidden dimension, and we use $h_i^{(l)}\!\in\!\mathbb{R}^{d}$ (i.e., the $i$-th row of $H^{(l)}$) to denote the hidden feature vector for the $i$-th gene.
In practice, $m_g^{(l)}$ can be implemented using standard message-passing architectures, such as GCN~\cite{gcn}, GraphSAGE~\cite{graphsage}, or GAT~\cite{gats}.
With increasing layer depth, the representations $H^{(l)}$ progressively aggregate broader neighborhood information, thereby capturing higher-order global regulatory interactions.

\subsubsection{Pairwise Branch}
To capture the highly specific, localized regulatory mechanisms governing individual TF–target pairs $(g_i, g_j)$, we design a dual-pathway branch consisting of an expression-based Transformer module and a structure-aware Subgraph module.

\noindent\textbf{Transformer Module.} Building upon the pairwise learning paradigm, we first extract the cell-wise expression profiles for the candidate pair, denoted as $e_i \!=\! X_{:,i}$ and $e_j \!=\! X_{:,j}$. To capture complex, non-linear co-expression patterns across the cell population, these raw features are projected via an MLP. They are then summed with learnable binary positional embeddings to explicitly distinguish their regulatory roles (i.e., TF versus target gene) and processed by a Transformer encoder:
\begin{equation}
z_{tr} = m_{tr}(e_i, e_j),
\end{equation}
where $m_{tr}$ denotes the Transformer-based encoding model, and $z_{tr}$ represents the fine-grained co-expression features.

\noindent\textbf{Subgraph Module.}
Gene regulation is rarely an isolated pairwise event; it frequently relies on the structural context provided by intermediate regulators or co-factors. We explicitly model these local micro-environments through a dynamic subgraph learning mechanism. 
Rather than relying on standard $k$-hop expansion, which often introduces noisy or irrelevant nodes, we adopt a priority-based subgraph expansion strategy to construct a local subgraph centered on the target pair $(g_i, g_j)$.
This design is motivated by the biological observation that regulatory interactions are often mediated by a subset of genes that are both topologically influential and functionally correlated.

At each expansion step, for every candidate neighbor $v$, we compute a hybrid importance score $S(v)$:
\begin{equation}
\label{eq:subgraph_score}
S(v) = w_{d} \cdot \underbrace{\frac{d(v)}{d_{max}}}_{\text{Topology}} + w_{e} \cdot \underbrace{\max\big(\rho(X_{:,v}, X_{:,i}),\, \rho(X_{:,v}, X_{:,j})\big)}_{\text{Expression}},
\end{equation}
where $d(v)$ is the node degree (reflecting its potential as a regulatory hub), $d_{max}$ is the maximum degree in the global graph, and $\rho(\cdot, \cdot)$ denotes the cosine similarity between gene expression profiles. The two terms jointly capture topological importance and functional co-expression relevance, with their trade-off controlled by hyperparameters $w_d$ and $w_e$.

Guided by this scoring mechanism, we select the top-$k$ neighbors to construct a compact yet informative subgraph, represented by its adjacency matrix $A_{sub}$ and node feature matrix $X_{sub}$.
A lightweight GNN, denoted as $m_{s}$, is then applied to encode this local graph, and the embeddings of the central pair are extracted to form the structural context vector:
\begin{equation}
z_{sub} = \left[ m_{s}(X_{sub}, A_{sub})_i \parallel m_{s}(X_{sub}, A_{sub})_j \right],
\end{equation}
where $\parallel$ denotes the concatenation operation, and $m_{s}(\cdot)_i$ denotes extracting the vector of gene $g_i$ from the GNN output.

Finally, we integrate the expression features and the structural context via a linear projection to obtain the pairwise representation for the downstream fusion stage:
\begin{equation}
z^{(0)}_{ij} = W_p [z_{tr} \parallel z_{sub}] + b_p,
\end{equation}
where $W_p$ and $b_p$ are learnable projection parameters.

\subsubsection{Multi-Scale Fusion}
To fully exploit both expression-level dynamics and the global topological context, we design a hierarchical fusion module that injectsthe global embeddings $h_i^{(l)}$ and $h_j^{(l)}$ into the local pairwise representation. 
Unlike static, single-step feature concatenation, this formulation enables the iterative refinement of pairwise representations conditioned on increasingly higher-order structural contexts.
Starting from the initial pairwise representation, we perform the following fusion recursively:
\begin{equation}
z_{ij}^{(l)} = m_{f}^{(l)}\left( z_{ij}^{(l-1)},[ h_i^{(l)} \parallel h_j^{(l)} ] \right),
\end{equation}
where $m_f^{(l)}$ denotes the layer-specific fusion model, implemented as an identity mapping for direct concatenation at $l=1$, and as a learnable gated mechanism for adaptive integration at $l>1$.
This design reflects the hierarchical nature of gene regulation, with local interactions modulated by broader network context.

The final fused representation $z_{ij}^{(L)}$ is fed into a prediction head composed of MLP layers with residual connections and layer normalization for stable optimization.
In this work, we instantiate the general objective $\mathcal{L}_{\mathrm{GRN}}$ as a binary cross-entropy loss over all candidate pairs in the training set:
\begin{equation}
\mathcal{L}_{\mathrm{ours}} = - \sum_{(g_i, g_j)\in E_{\mathrm{tr}}}
\bigl[y_{ij}\log\hat{y}_{ij} + (1 - y_{ij})\log(1 - \hat{y}_{ij})\bigr],
\end{equation}
where $y_{ij} \in \{0,1\}$ denotes the binary ground-truth label, and $\hat{y}_{ij} = \sigma(\small{\text{MLP}}({z}_{ij}^{(L)}))$ represents the predicted probability for the regulatory interaction $(g_i, g_j)$ generated by the prediction head.

\section{Experimental Setup}

\subsection{Datasets and data preprocessing}
\noindent\textbf{Datasets.} We evaluate the performance of \textit{scGPL} on seven scRNA-seq datasets curated by the BEELINE framework~\cite{2020Benchmarking}, covering diverse human and mouse cell types: (i) human embryonic stem cells (hESC), (ii) human hepatocytes (hHEP), (iii) mouse embryonic stem cells (mESC), (iv) mouse dendritic cells (mDC), (v) mouse hematopoietic stem cells with erythroid lineage (mHSC-E), (vi) granulocyte-monocyte lineage (mHSC-GM), and (vii) lymphoid lineage (mHSC-L). 

Ground-truth GRNs are provided for each dataset, integrated from multiple sources of biological evidence, including STRING functional interaction networks~\cite{string}, non-specific~\cite{nonspecific2,nonspecific1} and cell-type-specific ChIP-seq networks~\cite{Xu2013ESCAPE,specific2}, and LOF/GOF perturbation-based networks~\cite{Xu2013ESCAPE}. 
All datasets are publicly available from GEO: GSE75748 (hESC), GSE81252 (hHEP), GSE98664 (mESC), GSE48968 (mDC), and GSE81682 (mHSC variants).
Each scRNA-seq dataset comprises a varying number of transcription factors combined with either the top 500 or 1000 highly variable genes (denoted as TFs+500 and TFs+1000).

\noindent\textbf{Data Preprocessing.} We follow the preprocessing and splitting protocol in~\cite{genelink}, retaining only TF–target interactions and filtering genes by expression variance ($p < 0.01$, Bonferroni-corrected) as in~\cite{2020Benchmarking}.
Positive samples are defined as edges in the ground-truth networks, while all other TF–target pairs are treated as negative samples.
Given the sparsity of true regulatory networks, negative samples vastly outnumber positives, and some may correspond to undiscovered regulations~\cite{dataspilt1}.
We employ a TF-wise stratified splitting strategy. For TFs with more than two positive samples, their associated pairs are partitioned into $67\%$ training, $10\%$ validation, and $23\%$ testing sets, where the proportion of positives approximately matches the density of the underlying ground-truth regulatory network.
Conversely, for TFs with two or fewer positive samples, their associated pairs are distributed to preserve representation: TFs with exactly two samples are evenly split between the training and testing sets, while those with a single sample are randomly assigned to either set. 
Crucially, edges in the test set are strictly removed from the training graph and are not accessible during subgraph construction or message passing, preventing information leakage. Detailed dataset statistics are provided in Supplementary Tables 1--2

\subsection{Baselines and Evaluation Metrics}

\noindent\textbf{Baseline Methods.} We compare \textit{scGPL} against thirteen representative GRN inference methods: GENIE3~\cite{genie3}, mutual information-based method (MI)~\cite{song2012mi}, GRNBoost2~\cite{moerman2019grnboost2}, Pearson correlation coefficient (PCC), GNE, DeepSEM~\cite{shu2021deepsem}, GENELink, GNNLink, scGREAT, as well as recent methods including scRegNet~\cite{kommu2025prediction}, GT-GRN~\cite{Teji2025GTGRN}, GRLGRN, and GRNFormer.
Among these, GENELink and GNNLink are representative graph-based methods, while GNE and scGREAT follow the pairwise learning paradigm.
All models are evaluated on the same datasets and splits, with hyperparameters configured according to their original implementations when available, to ensure a fair comparison.

\noindent\textbf{Evaluation Metrics.}
We evaluate our model using three metrics: Area Under the Precision–Recall Curve (AUPRC), Area Under the Receiver Operating Characteristic Curve (AUROC), and Early Precision Ratio (EPR).
Given the severe class imbalance in GRN datasets, AUPRC serves as our primary metric owing to its sensitivity to false positives, which precisely reflects performance on sparse true edges.
AUROC is utilized as a complementary measure to reflect overall ranking capability and structural robustness.
Additionally, EPR measures the density ratio of true positive interactions in the top-$k$ predictions relative to a random expectation, with $k=100$ following \cite{cui2025infosemdeepgenerativemodel}.
Metric definitions are provided in Supplementary D.1.

\begin{table*}[!t]
\centering

\caption{Performance comparison of GRN inference methods across seven datasets using cell-type-specific ChIP-seq networks. Results are reported for both TFs+500 and TFs+1000 settings in terms of AUROC, AUPRC, and EPR. All baseline methods were faithfully reproduced under identical experimental settings. The reported results represent the average of 20 independent runs using different random seeds. The best and second-best results in each row are highlighted in \textbf{bold} and \underline{underlined}, respectively.}
\label{tab:performance_comparison}
\resizebox{\textwidth}{!}{
\begin{tabular}{c | c ccc ccc ccc ccc ccc}
\toprule
\multirow{2}{*}{\textbf{Dataset}} & \multirow{2}{*}{\textbf{Cell Type}} & \multicolumn{3}{c}{\textbf{GNNLINK}} & \multicolumn{3}{c}{\textbf{GENELINK}} & \multicolumn{3}{c}{\textbf{scRegNet}} & \multicolumn{3}{c}{\textbf{scGREAT}} & \multicolumn{3}{c}{\textbf{scGPL}} \\
\cmidrule(lr){3-5} \cmidrule(lr){6-8} \cmidrule(lr){9-11} \cmidrule(lr){12-14} \cmidrule(lr){15-17}
& & AUROC & AUPRC & EPR & AUROC & AUPRC & EPR & AUROC & AUPRC & EPR & AUROC & AUPRC & EPR & AUROC & AUPRC & EPR \\
\midrule
\multirow{8}{*}{\rotatebox{90}{TFs+500}} 
& hESC    & 0.821 & 0.419 & 3.288 & 0.789 & 0.455 & 5.464 & 0.874 & 0.607 & \underline{6.328} & \underline{0.888} & \underline{0.632} & 6.218 & \textbf{0.917} & \textbf{0.728} & \textbf{6.360} \\
& hHEP    & 0.741 & 0.546 & 1.832 & 0.845 & 0.707 & 2.434 & 0.896 & 0.833 & \underline{2.839} & \underline{0.909} & \underline{0.856} & 2.828 & \textbf{0.925} & \textbf{0.875} & \textbf{2.842} \\
& mDC     & 0.771 & 0.126 & 3.112 & 0.726 & 0.123 & 3.607 & \underline{0.813} & \underline{0.173} & \underline{5.654} & 0.802 & 0.148 & 3.957 & \textbf{0.890} & \textbf{0.427} & \textbf{11.665}\\
& mESC    & 0.774 & 0.542 & 2.956 & 0.878 & 0.747 & 2.820 & 0.924 & 0.855 & \textbf{3.318} & \underline{0.935} & \underline{0.877} & \textbf{3.318} & \textbf{0.945} & \textbf{0.891} & \textbf{3.318} \\
& mHSC-E  & 0.735 & 0.716 & 1.370 & 0.874 & 0.889 & 1.702 & 0.918 & 0.937 & \underline{1.762} & \underline{0.928} & \underline{0.947} & \textbf{1.766} & \textbf{0.936} & \textbf{0.951} & \textbf{1.766} \\
& mHSC-GM & 0.789 & 0.732 & 1.252 & 0.891 & 0.895 & 1.868 & \underline{0.935} & 0.942 & \underline{1.880} & 0.934 & \underline{0.944} & 1.874 & \textbf{0.940} & \textbf{0.947} & \textbf{1.885} \\
& mHSC-L  & 0.727 & 0.654 & 1.214 & 0.829 & 0.826 & 1.950 & \underline{0.884} & 0.883 & \underline{2.018} & 0.881 & \underline{0.885} & \textbf{2.033} & \textbf{0.899} & \textbf{0.898} & \textbf{2.033} \\
\cmidrule{2-17}
& Avg.    & 0.765 & 0.534 & 2.146 & 0.833 & 0.663 & 2.835 & 0.892 & 0.747 & \underline{3.400} & \underline{0.897} & \underline{0.755} & 3.142 & \textbf{0.922} & \textbf{0.817} & \textbf{4.267} \\
\midrule
\multirow{8}{*}{\rotatebox{90}{TFs+1000}} 
& hESC    & 0.818 & 0.393 & 2.180 & 0.823 & 0.483 & 5.108 & 0.881 & 0.620 & 6.345 & \underline{0.890} & \underline{0.641} & \underline{6.493} & \textbf{0.931} & \textbf{0.786} & \textbf{6.542} \\
& hHEP    & 0.762 & 0.559 & 1.600 & 0.844 & 0.707 & 2.413 & 0.901 & 0.843 & \underline{2.880} & \underline{0.911} & \underline{0.857} & \underline{2.880} & \textbf{0.937} & \textbf{0.890} & \textbf{2.890} \\
& mDC     & 0.782 & 0.148 & 5.443 & 0.755 & 0.123 & 3.470 & \underline{0.838} & \underline{0.168} & \underline{6.987} & 0.814 & 0.149 & 5.511 & \textbf{0.943} & \textbf{0.671} & \textbf{19.654}\\
& mESC    & 0.775 & 0.539 & 2.981 & 0.890 & 0.769 & 2.927 & 0.933 & 0.868 & \textbf{3.420} & \underline{0.942} & \underline{0.888} & 3.360 & \textbf{0.950} & \textbf{0.897} & \underline{3.372} \\
& mHSC-E  & 0.770 & 0.737 & 1.505 & 0.895 & 0.906 & 1.785 & 0.940 & 0.951 & \underline{1.788} & \underline{0.942} & \underline{0.956} & \textbf{1.807} & \textbf{0.950} & \textbf{0.959} & \textbf{1.807} \\
& mHSC-GM & 0.797 & 0.749 & 1.265 & 0.902 & 0.913 & 1.817 & 0.842 & 0.949 & \underline{1.876} & \underline{0.942} & \underline{0.953} & 1.841 & \textbf{0.951} & \textbf{0.958} & \textbf{1.895} \\
& mHSC-L  & 0.751 & 0.659 & 1.319 & 0.838 & 0.815 & 2.043 & \underline{0.884} & \underline{0.875} & 2.116 & 0.883 & 0.873 & \underline{2.118} & \textbf{0.900} & \textbf{0.885} & \textbf{2.123} \\
\cmidrule{2-17}
& Avg.    & 0.779 & 0.541 & 2.328 & 0.850 & 0.674 & 2.795 & 0.888 & 0.753 & \underline{3.630} & \underline{0.903} & \underline{0.760} & 3.430 & \textbf{0.937} & \textbf{0.864} & \textbf{5.469} \\
\bottomrule
\end{tabular}
}
\end{table*}

\subsection{Implementation Details}All models are trained on eight NVIDIA RTX 4090 GPUs. The Global Branch employs a two-layer GAT with 4 heads, projecting inputs to 512-dimensional embeddings and subsequently compressing them to 256 dimensions.
Within the Pairwise Branch, the Subgraph Module similarly utilizes a two-layer GAT with 4 heads, mapping subgraph inputs to 256 dimensions followed by 128 dimensions, with the expansion weights set to $w_d=0.4$ and $w_e=0.6$, a subgraph hop number of 2, and a maximum node capacity of 100.
Meanwhile, the Transformer Module models feature interactions via a four-layer Transformer encoder with 8 attention heads and an embedding size of 1024.

Optimization is performed using the Adam optimizer with a batch size of 256, a learning rate and weight decay of $1 \times 10^{-5}$, coupled with a step scheduler ($\gamma=0.999$, step size=10). The model is trained for up to 200 epochs, employing early stopping based on validation AUROC with a patience of 8 epochs. \textit{scGPL} is trained in an end-to-end manner, allowing gradients to propagate simultaneously to both global and local branches. Further details on the prediction head and hyperparameters are provided in Supplementary Material Table S3.

\section{Results}

\subsection{Performance on benchmark datasets}

\noindent\textbf{Overall Performance.}
As highlighted in Table~\ref{tab:performance_comparison}, \textit{scGPL} secures state-of-the-art results on all seven datasets under the cell-type-specific (hereafter referred to as Specific) networks. 
In terms of AUROC, \textit{scGPL} achieves the highest overall average, surpassing \textit{scGREAT} by 3.4\% on TFs+500 and further increasing this margin to a 3.8\% advantage on the larger TFs+1000 setting. This advantage is even more pronounced in AUPRC, where \textit{scGPL} improves upon \textit{scGREAT} by 13.7\% on TFs+1000.
Notably, the EPR score of \textit{scGPL} achieves a nearly three-fold improvement over \textit{scGREAT} on the mDC dataset, which features the fewest positive samples and the sparsest network density among the Specific networks. This demonstrates a superior capability in prioritizing true regulatory links among top-ranked predictions, consistent with its robust AUPRC performance. 
Furthermore, while the recent subgraph-based method GRNFormer achieves highly competitive performance on these networks, particularly excelling on the mESC and mHSC-L datasets (Supplementary Table S4), \textit{scGPL} still achieves higher AUROC and AUPRC by 1.2\% and 1.7\%, respectively.

\begin{figure}[t] 
    \centering
    \includegraphics[width=0.9\linewidth]{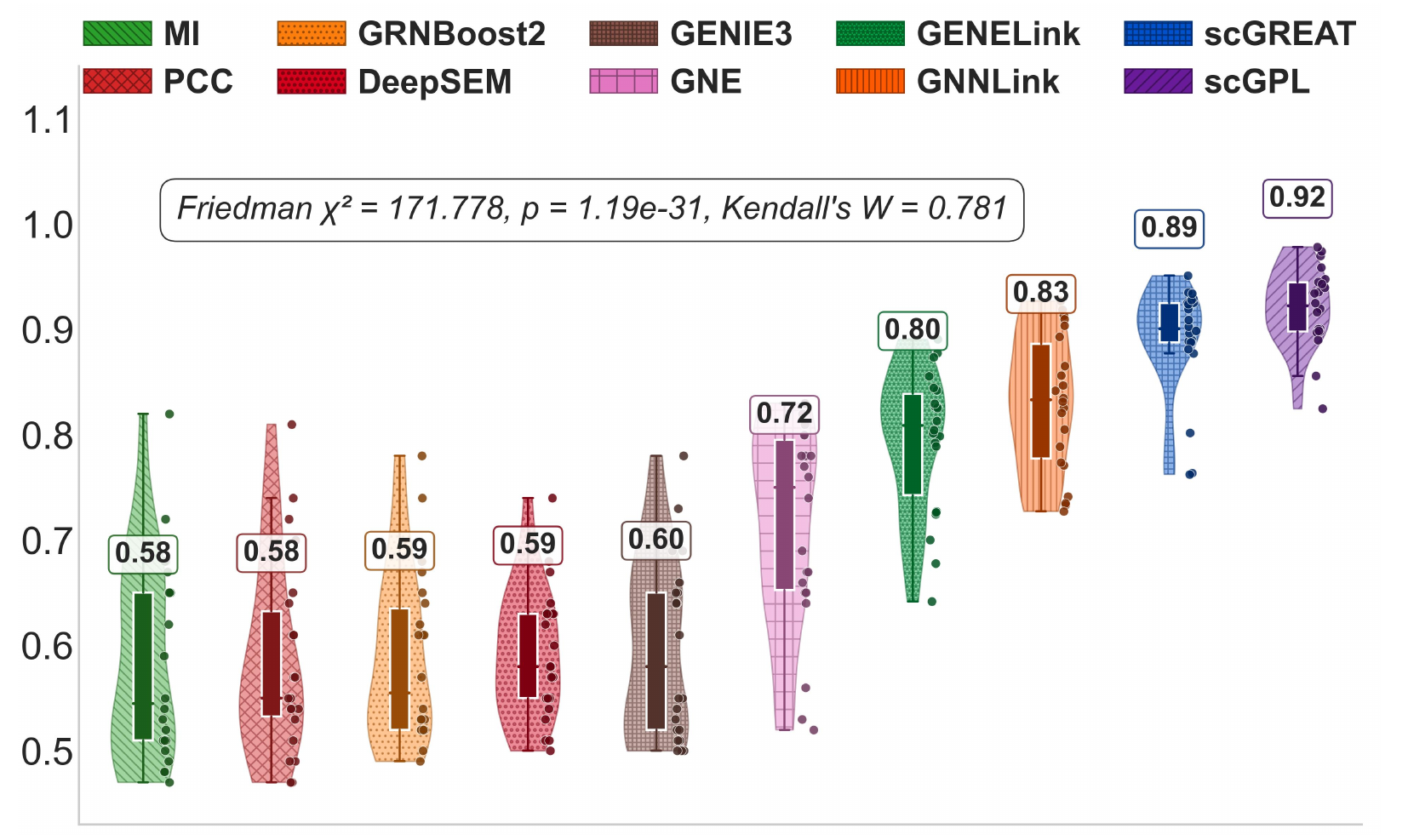}
    \caption{Raincloud plots illustrating the distribution of AUROC scores on TFs+500. Each plot combines a violin density estimate to show distribution shape, a central box plot (in darker shade) indicating the median and quartiles, and jittered scatter points plotted to the right of the box plot to represent individual dataset performance.}
    \label{fig3:auroc500raincloud}
\end{figure}

\begin{figure}[t] 
    \centering
    \includegraphics[width=0.9\linewidth]{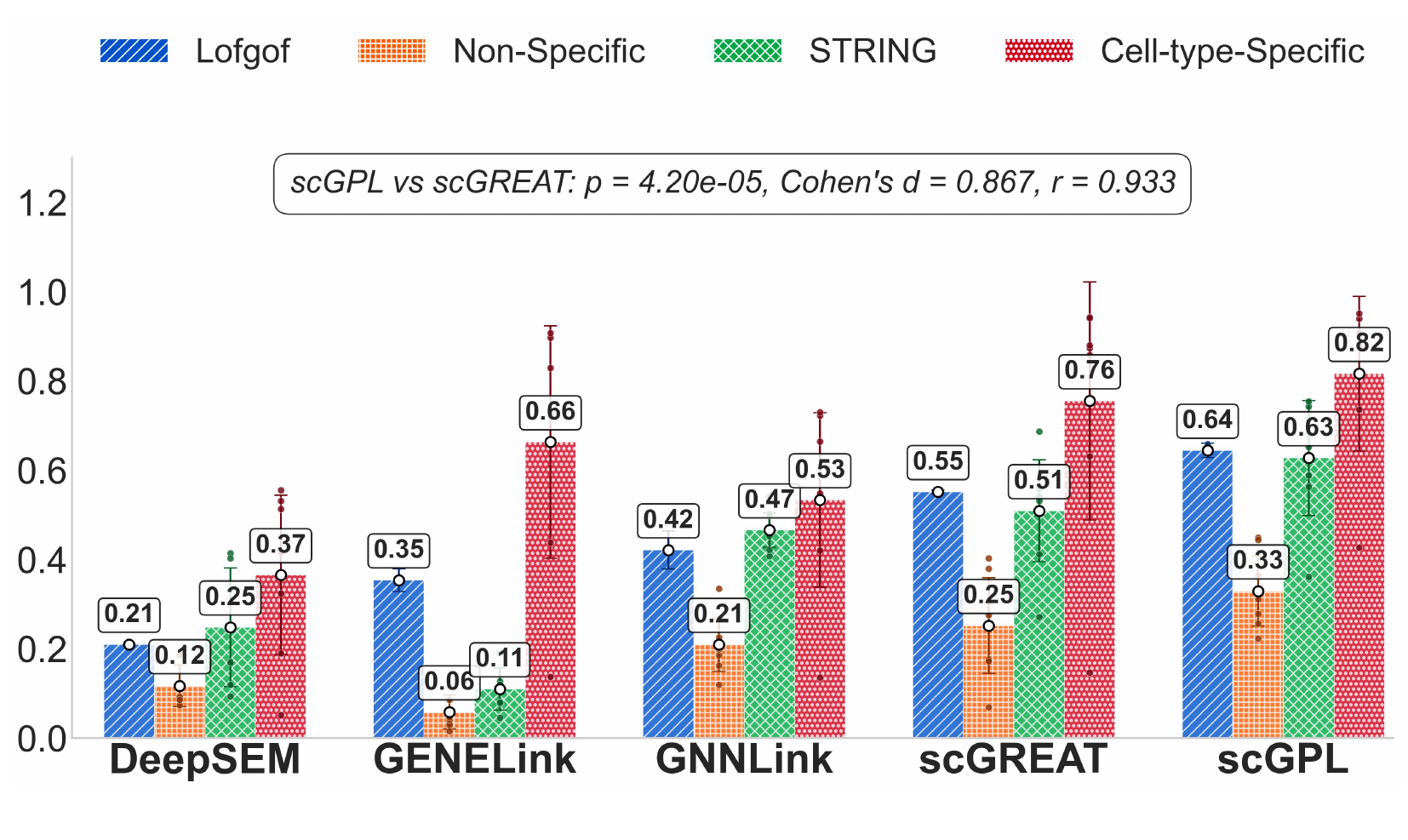} 
    \caption{AUPRC performance of five GRN inference methods under four networks. Each method is evaluated under four distinct ground truth networks on TFs+500: STRING (green), Non-Specific (orange), LOF/GOF (blue), and cell-type-Specific (red). Error bars denote standard deviation across different cell types.}
    \label{fig4:auprc500bar}
\end{figure}

Beyond the Specific networks, \textit{scGPL} consistently delivers outstanding performance across all 42 evaluated benchmarks, as detailed in the heatmaps (Figures~S1--S3). 
Figure~\ref{fig3:auroc500raincloud} presents the AUROC distributions on TFs+500, highlighting the clear advantage of \textit{scGPL} over other methods.
The performance gains are especially pronounced on highly imbalanced and sparse networks, such as the STRING and Non-Specific networks (Figure~\ref{fig4:auprc500bar}), where \textit{scGPL} achieves AUPRC improvements of approximately 23.5\% and 32.0\% over \textit{scGREAT}, respectively.
We note that the Graph Transformer-based method GRLGRN emerges as the strongest baseline on the STRING networks under the TFs+500 setting (Table~S4), but \textit{scGPL} maintains its superiority under both high-signal conditions and sparse networks (Figures~S4--S8).
Moreover, the overall AUROC and AUPRC (Figures~S9--S12) confirm \textit{scGPL}'s superiority, while EPR evaluations (Figures~S13--S14) report average scores of 32.04 and 44.99 on TFs+500 and TFs+1000, respectively, outperforming \textit{scGREAT} (28.19, 36.87) and \textit{GNNLink} (25.50, 31.66).
This further confirms its exceptional capacity to prioritize true positive regulatory edges across diverse networks.

\noindent\textbf{Statistical Significance.} 
To rigorously assess the significance of \textit{scGPL}'s performance gains, we evaluated 10 GRN inference methods on the TFs+500 datasets across 22 cell types under four types of ground-truth networks. Friedman tests reveal highly significant differences among methods for both AUROC (annotated on Figure~\ref{fig3:auroc500raincloud}) and AUPRC ($\chi^2=129.340$, $p=6.37\times10^{-23}$, Kendall's $W=0.588$). Furthermore, Nemenyi post-hoc tests confirm that \textit{scGPL} consistently ranks first and differs significantly from most baselines. To further validate these findings, we performed paired tests against the strongest baseline, \textit{scGREAT}, showing that \textit{scGPL} achieves significantly higher AUROC (Wilcoxon $p \approx 1.43\times10^{-6}$, Cohen's $d \approx 1.08$, Pearson $r \approx 0.87$) and AUPRC (Figure~\ref{fig4:auprc500bar}), with particularly strong gains on STRING and Non-Specific networks. 
Additional significance test results, visualizations, and analyses are provided in the Supplementary Figures~S4--S14 and Section~D.2.

\subsection{Ablation study}

\begin{table}[!t]
\centering
\caption{Ablation study of the Pairwise branch and Global Branch in our integral design: AUPRC comparison under Specific and STRING networks on TFs+500.}
\label{tab:ablation_study_1}
\resizebox{\columnwidth}{!}{
\begin{tabular}{@{}l cccccc @{}}

\toprule
\multirow{3}{*}{\textbf{Dataset}} 
& \multicolumn{2}{c}{\textbf{Pairwise Only}} 
& \multicolumn{2}{c}{\textbf{Global Only}} 
& \multicolumn{2}{c}{\textbf{Both Branches}} \\
\cmidrule(lr){2-3} \cmidrule(lr){4-5} \cmidrule(lr){6-7}
& \textbf{STRING} & \textbf{Specific} 
& \textbf{STRING} & \textbf{Specific} 
& \textbf{STRING} & \textbf{Specific} \\
\midrule
\textbf{hESC}     & \textbf{0.617} & \underline{0.619} & 0.281 & 0.456 & \underline{0.561} & \textbf{0.654} \\
\textbf{hHEP}     & \underline{0.681} & \underline{0.853} & 0.322 & 0.707 & \textbf{0.713} & \textbf{0.866} \\
\textbf{mDC}      & \textbf{0.747} & \underline{0.136} & 0.431 & 0.123 & \underline{0.712} & \textbf{0.453} \\
\textbf{mESC}     & \underline{0.625} & \textbf{0.888} & 0.188 & 0.747 & \textbf{0.766} & \underline{0.853} \\
\textbf{mHSC-E}   & 0.182 & \underline{0.943} & \underline{0.303} & 0.889 & \textbf{0.547} & \textbf{0.944} \\
\textbf{mHSC-GM}  & 0.357 & \textbf{0.941} & \underline{0.365} & 0.895 & \textbf{0.585} & \underline{0.935} \\
\textbf{mHSC-L}   & 0.090 & \textbf{0.883} & \underline{0.130} & 0.827 & \textbf{0.326} & \underline{0.880} \\
\midrule
\textbf{Average}  & \underline{0.471} & \underline{0.752} & 0.289 & 0.663 & \textbf{0.602} & \textbf{0.798} \\
\bottomrule
\end{tabular}
}
\end{table}
\begin{table}[!t]
\centering
\caption{Ablation study of the Subgraph Module in our integral design: AUPRC comparison under two networks on TFs+500.}
\label{tab:ablation_study_2}
\resizebox{\columnwidth}{!}{
\begin{tabular}{@{}l cccccc @{}}
\toprule
\multirow{3}{*}{\textbf{Dataset}} 
& \multicolumn{2}{c}{\textbf{w/o Subgraph Module}} 
& \multicolumn{2}{c}{\textbf{w/ Subgraph Module}} 
& \multicolumn{2}{c}{\textbf{w/ Subgraph Module}} \\
& \multicolumn{2}{c}{} 
& \multicolumn{2}{c}{\textbf{(w/BFS)}} 
& \multicolumn{2}{c}{\textbf{(w/Priority)}} \\
\cmidrule(lr){2-3} \cmidrule(lr){4-5} \cmidrule(lr){6-7}
 & \textbf{STRING} & \textbf{Specific} 
 & \textbf{STRING} & \textbf{Specific} 
 & \textbf{STRING} & \textbf{Specific} \\
\midrule
\textbf{hESC}     & 0.561 & 0.654 & \underline{0.628} & \underline{0.705} & \textbf{0.661} & \textbf{0.728} \\
\textbf{hHEP}     & 0.713 & 0.866 & \underline{0.717} & \underline{0.872} & \textbf{0.730} & \textbf{0.875} \\
\textbf{mDC}      & 0.712 & \textbf{0.453} & \underline{0.713} & 0.423 & \textbf{0.724} & \underline{0.427} \\
\textbf{mESC}     & \textbf{0.766} & 0.853 & 0.749 & \underline{0.880} & \textbf{0.766} & \textbf{0.891} \\
\textbf{mHSC-E}   & \underline{0.547} & 0.944 & 0.539 & \underline{0.950} & \textbf{0.558} & \textbf{0.951} \\
\textbf{mHSC-GM}  & \underline{0.585} & \underline{0.935} & 0.571 & 0.927 & \textbf{0.605} & \textbf{0.947} \\
\textbf{mHSC-L}   & 0.326 & 0.880 & \underline{0.342} & \underline{0.898} & \textbf{0.354} & \textbf{0.898} \\
\midrule
\textbf{Average}  & 0.602 & 0.798 & \underline{0.608} & \underline{0.808} & \textbf{0.628} & \textbf{0.817} \\
\bottomrule
\end{tabular}
}
\end{table}

To elucidate the contributions of key components in \textit{scGPL}, we conducted ablation using two networks with distinct densities.

\noindent\textbf{Dual-branch.} 
First, we evaluate the effectiveness of the dual-stream architecture by comparing three variants: (i) the Pairwise Only branch (utilizing only the Transformer), (ii) the Global Only branch (utilizing the GAT), and (iii) the integrated Dual-Branch model (combining the pairwise Transformer with the global branch).
As shown in Table~\ref{tab:ablation_study_1}, on the relatively balanced Specific networks, the Pairwise Only branch alone performs competitively with an average AUPRC of 0.7519, surpassing the Global Only branch (0.6633).
However, the integrated model achieves a significant improvement, reaching an average AUPRC of 0.7978.
On the highly imbalanced STRING networks, the advantages of the integrated model are even more pronounced, outperforming the Pairwise Only and Global Only variants by 27.7\% and 108.49\%, respectively.
This demonstrates that global topological information provides indispensable auxiliary features, effectively correcting the limitations of a single local perspective.

\noindent\textbf{Subgraph module.} 
Next, building upon the dual-branch architecture, we evaluate the specific impact of the Subgraph Module. We treat the integrated model without subgraph input as the baseline (denoted as "w/o Subgraph Module") and compare it against two extraction strategies: standard BFS ("w/ Subgraph Module (w/BFS)") and our proposed method ("w/ Subgraph Module (w/Priority)"). As presented in Table~\ref{tab:ablation_study_2}, the priority-based iterative expansion strategy yields the best performance. While the standard BFS offers only marginal gains, the priority-based module achieves substantial improvements, boosting performance by 4.42\% on STRING.
Furthermore, to validate the design of \textit{scGPL} localized context extraction, we compared our pair-specific against a TF-centric sampling strategy.
As shown in Supplementary Table~S6, the pair-specific strategy consistently outperforms the TF-centric method across networks with varying densities, achieving a higher average AUPRC on the STRING networks (0.628 versus 0.626) and further widening the performance gap on the Specific networks (0.817 versus 0.808).
This indicates that our algorithm can effectively capture the most informative local topologies tailored to specific regulatory interactions.
Additionally, evaluations across different GNN backbones (Supplementary Figure~S15 and Section D.3) confirm that \textit{scGPL} maintains competitive predictive performance regardless of the specific graph encoder variant used.
Collectively, these results further confirm that both the dual-stream architecture and dynamic subgraph extraction are crucial for the superior performance of \textit{scGPL}.



\subsection{Computational Efficiency}

\begin{table}[!t]
\centering
\caption{Comparison of computational efficiency across all methods on the hESC dataset under Specific on TFs+500.}
\label{tab:method_comparison}
\resizebox{\columnwidth}{!}{%
\begin{tabular}{l cccc}
\toprule
\textbf{Method} & \textbf{Time/Epoch (s)} & \textbf{Avg Mem (MB)} & \textbf{Total Time} & \textbf{Params (M)} \\
\midrule
DeepSEM   & $0.13 \pm 0.01$  & $7.08 \pm 0.00$   & 12.69 s       & 0.83  \\
CNNC      & $30.55 \pm 0.14$  & $1300.31 \pm 3.48$  & 48 min 7s      & 398.22  \\
GNE       & $18.58 \pm 0.11$  & $324.16 \pm 0.98$   & 23 min 39s        & 92.14  \\
GENELink  & $2.53 \pm 0.28$  & $78.04 \pm 0.19$  & 2 min 6s      & 1.74  \\
scGREAT   & $4.81 \pm 0.89$  & $258.81 \pm 0.08$ & 14 min 32s      & 20.26 \\
scGPL     & $14.99 \pm 6.22$ & $699.86 \pm 0.57$ & 11 min 06s    & 50.59 \\

\bottomrule
\end{tabular}%
}
\end{table}

To comprehensively evaluate the computational efficiency of \textit{scGPL}, Table~\ref{tab:method_comparison} compares its training costs against evaluated baselines. Although the per-epoch training time of \textit{scGPL} is higher than that of \textit{GENELink} and \textit{scGREAT}, evaluating per-epoch cost alone is insufficient for an end-to-end efficiency comparison, as the total epochs required for convergence significantly impact the overall cost. Ultimately, the total training time for \textit{scGPL} is merely 666 seconds, which is faster than \textit{scGREAT} and substantially lower than \textit{GNE} and \textit{CNNC} (Figure~S16a, d). Figure~S16b further illustrates this dynamic: \textit{scGPL} converges in just 44 epochs, whereas \textit{scGREAT} requires nearly 160 epochs, and \textit{GENELink} stops at under 20 epochs. This rapid convergence demonstrates that \textit{scGPL} successfully inherits the efficient convergence advantages of graph-based methods; despite the lower time costs of pure GNN baselines, \textit{scGPL} surpasses the performance of pairwise-based methods and achieves the highest AUPRC among all models (Figure~S16c).
Beyond training time, we also evaluated inference latency and resource overhead. \textit{scGPL} contains 50.59 million parameters, with average and peak training GPU memory consumption of 699.86 MB and 1062.57 MB, respectively, remaining lower than those of \textit{GNE} and \textit{CNNC} (Figure~S17a, b). During the inference phase, \textit{scGPL} requires a virtually negligible average time of 3.99 seconds and a memory footprint of 702.49 MB (Figure~S17c, d). Although its latency and memory efficiency may be slightly inferior to the most lightweight baselines, its state-of-the-art predictive performance across all datasets fully justifies these acceptable computational trade-offs.

\subsection{Sensitivity and Scalability Analysis}
We further conducted a hyperparameter sensitivity and scalability analysis to evaluate the impact of batch size, subgraph configuration, and the scale of genes and cells on both computational overhead and predictive performance.

\noindent\textbf{Impact of Batch Size.}
Increasing the batch size leads to a slight degradation in validation performance and a continuous rise in peak GPU memory usage, as illustrated in Figure~S18. Interestingly, training time does not scale linearly with batch size.
Specifically, training times at batch sizes of 32 and 64 exceed those at 128, 256, 512, and 1024, while training throughput reaches its maximum at 128, rising below this point and decreasing above it.
This phenomenon is likely attributable to the computational overhead of irregular batching with heterogeneous subgraphs (varying node counts per sample), which diminishes GPU memory access efficiency beyond the critical threshold, thereby degrading overall performance.

\noindent\textbf{Impact of Subgraph Parameters.}
As shown in Figure~S19, the maximum number of nodes per subgraph is the primary determinant of both predictive performance and computational cost, whereas the number of hops exerts a marginal effect. Increasing the subgraph size steadily improves validation performance, but at the cost of significantly higher per-epoch training time and peak GPU memory. In contrast, increasing the number of hops under a fixed subgraph size yields negligible performance differences, indicating that retaining more informative nodes is more effective than expanding the receptive field depth. A highly practical trade-off is achieved at a subgraph size of 100 with two hops, which captures over 99\% of the optimal predictive performance while maintaining low time and memory overheads. For new datasets, a subgraph size of 20--30 serves as a cost-effective initial range, achieving better hardware efficiency while maintaining robust signal capture.

\noindent\textbf{Impact of Gene and Cell Numbers.}
To evaluate the scalability of \textit{scGPL} across different data scales, we performed controlled truncation on TFs+1000, dividing cells into four gradients (25\%, 50\%, 75\%, and 100\%) and genes into five gradients by incorporating the TFs+500 setting as an additional dimension.
We observed that for the hESC dataset, as the number of genes increases, the predictive performance slightly improves on the STRING network (Figure~S21a), but exhibits a modest decline on the Specific network (Figure~S22a).
Regarding the number of cells, performance consistently improves with increasing cell numbers across both networks, demonstrating the ability of \textit{scGPL} to effectively exploit expression-derived signals (Figures~S21d and S22d).
In terms of computational efficiency, both per-epoch training time and peak GPU memory usage generally rise with cell numbers, though total training time and memory costs fluctuate non-monotonically due to varying convergence epochs and subgraph extraction differences across gene densities (Figures~S21b--c, S21e--f, S22b--c, and S22e--f).
Overall, these results indicate that increasing the number of cells yields consistent performance gains, whereas the impact of increasing the number of genes is closely related to network sparsity.
Further analysis of computational complexity with respect to graph size is detailed in Section D.5.








\subsection{Robustness Analysis}

\begin{figure}[t!]
    \centering
    \includegraphics[width=\linewidth]{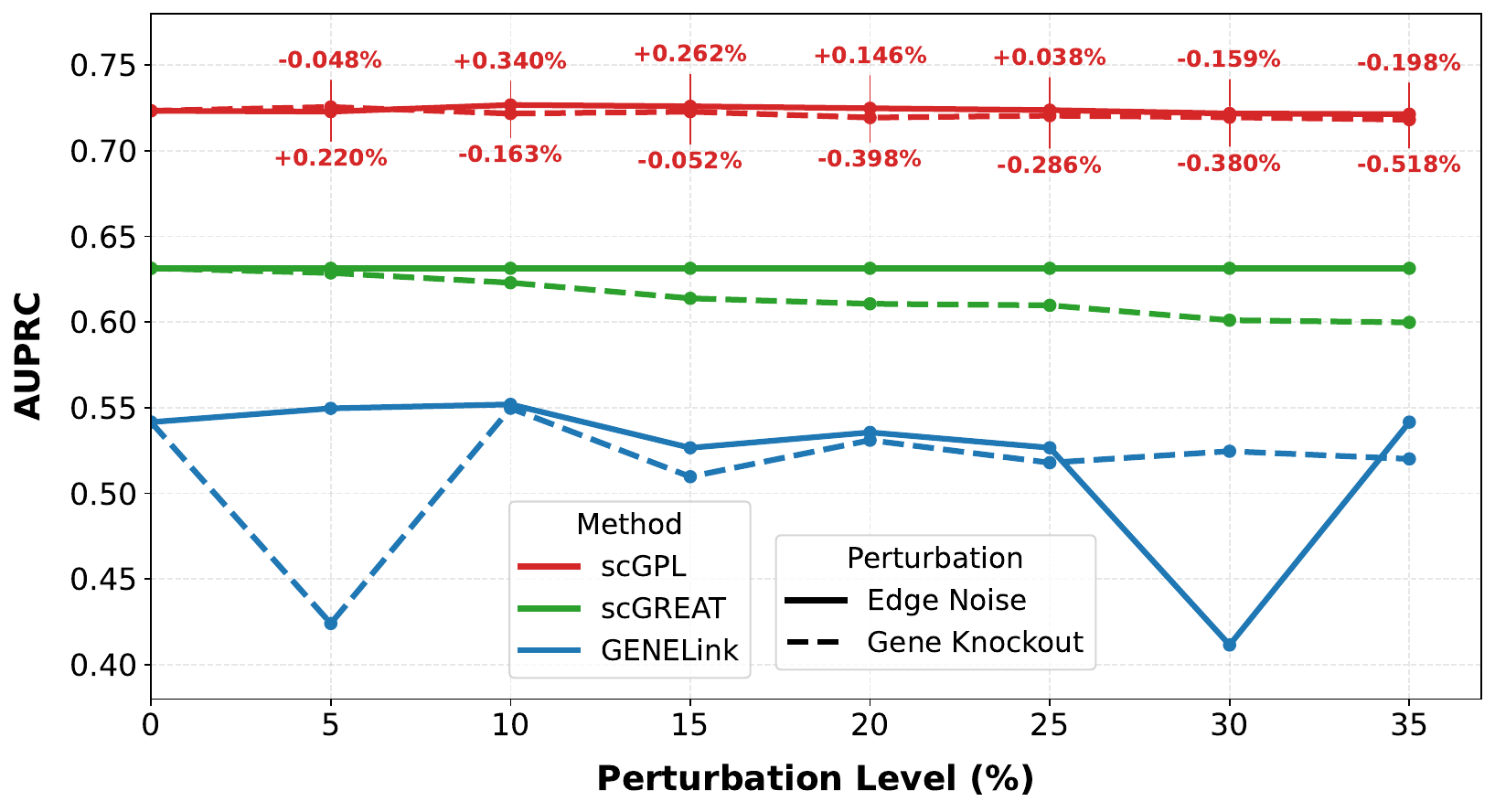}
    \caption{AUPRC plotted against perturbation intensity for Edge Noise (solid line) and Gene Knockout (dashed line). Annotations on the scGPL curves indicate the absolute deviation from the baseline AUPRC ($\Delta$AUPRC, in percentage points), positioned above the Edge Noise trace and below the Gene Knockout trace.}
    \label{fig6}
\end{figure}

Given the severe noise and sparsity inherent to real-world single-cell data, we assessed how stably \textit{scGPL} performs under such suboptimal conditions. To this end, we simulated perturbation scenarios of varying intensities (ranging from 0\% to 35\% with a 5\% step size): (i) Edge Noise, which injects random false-positive regulatory links; (ii) Edge Dropout, which randomly removes existing true links to simulate incomplete prior topological knowledge; and (iii) Gene Knockout, which masks the expression profiles of specific genes to zero vectors.
As depicted by the performance decay curves in Figure~\ref{fig6} (full results are provided in Figures~S23--S25), \textit{scGPL} exhibits remarkable robustness. For instance, under 30\% Edge Noise, the AUPRC of \textit{scGPL} decreases by merely 0.159\%, significantly outperforming the baseline \textit{GENELink}, which suffers a 13\% decline.
This demonstrates that our model effectively overcomes the performance instability typically associated with GNN models under edge dropout conditions. Similarly, in the Gene Knockout scenario, \textit{scGPL} experiences only a 0.518\% performance loss, which is vastly superior to the 3.16\% degradation observed in \textit{scGREAT}. This indicates that the dual-branch architecture compensates for missing expression signals by leveraging a broader global topological context.
These findings confirm that \textit{scGPL} is highly robust against both topological perturbations and expression anomalies.

\subsection{Biological interpretability}

\begin{figure}[t!]
    \centering
    \includegraphics[width=\linewidth]{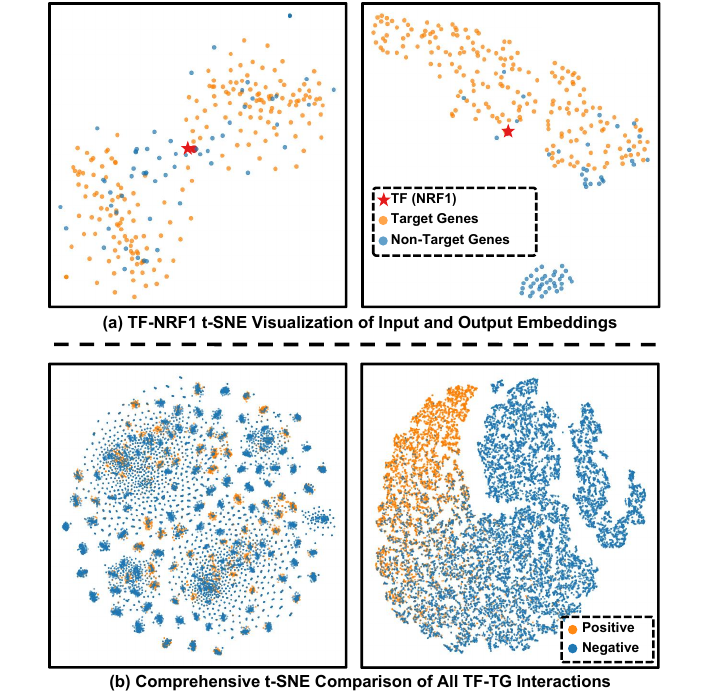}
    \caption{t-SNE visualization of transcription factor-target gene interactions. (a) Local view of TF \textit{NRF1} with target and non-target genes using raw input features (left) versus penultimate layer embeddings (right). (b) Global view of all TF-target pairs using raw input features (left) versus penultimate layer embeddings (right).}
    \label{fig7:tsne}
\end{figure}

\noindent\textbf{Representation Visualization.}
To evaluate the representation learning capability of \textit{scGPL}, we utilized t-SNE visualization to examine whether the model generates biologically meaningful and discriminative embeddings capable of distinguishing true TF-target gene pairs from non-interacting pairs.
First, we analyzed the TF \textit{NRF1} in the mESC dataset under the Specific network. Embeddings derived directly from raw expression profiles exhibited substantial overlap between target and non-target gene pairs, indicating poor separability in the original expression space. In contrast, the latent representations extracted from the penultimate layer of \textit{scGPL}—which seamlessly integrate gene expression, global topological properties, and local subgraph contexts—demonstrated distinct clustering. Specifically, \textit{NRF1} and its true target genes formed highly cohesive groups (Figure~\ref{fig7:tsne}(a)). This significant enhancement in intra-class cohesion and inter-class separation suggests that \textit{scGPL} successfully captures the synergistic regulatory mechanisms necessary for maintaining cell states.
Second, when scaling this analysis globally to all TF-target gene pairs, the \textit{scGPL} embeddings consistently maintained superior class separation relative to the raw input space (Figure~\ref{fig7:tsne}(b)). Collectively, these results demonstrate that \textit{scGPL} effectively transforms high-dimensional, sparse single-cell expression data into a robust feature space enriched with regulatory semantics. Additional TF-specific case studies and global embedding visualizations across other datasets are provided in Supplementary Figure~S26--S31.

\noindent\textbf{Case Study.}
While benchmark metrics effectively evaluate a model's ability to recover documented regulatory interactions, they are inherently limited by the incompleteness of current GRN annotations. 
A practical GRN inference framework should therefore not only reproduce known regulatory relationships but also provide biologically meaningful hypotheses beyond existing databases. 
To investigate this capability, we analyzed the top-ranked, high-confidence predictions generated by \textit{scGPL} that were absent from the benchmark ground-truth networks. 
Rather than dismissing these unannotated predictions as incorrect, we systematically examined them as potential missing regulatory relationships to determine whether our model could recover biologically plausible associations beyond current annotations.

\begin{figure}[t!]
    \centering
    \includegraphics[width=\linewidth]{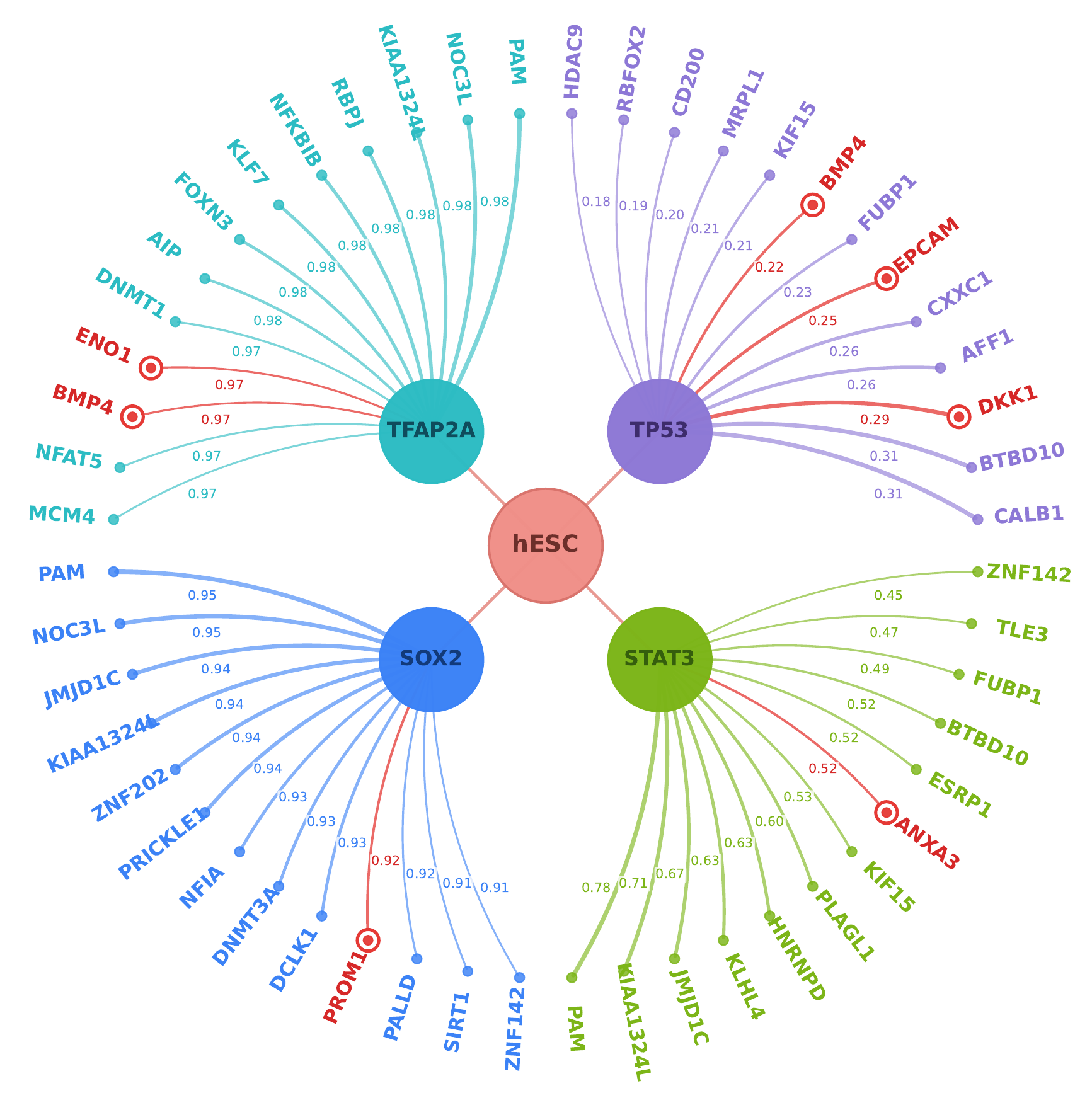}
    \caption{Visualization of the predicted regulatory network on the hESC dataset under the Specific network on TFs+500. Central, intermediate, and outer nodes represent the dataset, TFs, and predicted target genes, respectively.  Literature-supported targets are highlighted in red, with prediction scores annotated on the edges.}
    \label{fig8:grn}
\end{figure}

As illustrated in Figure~\ref{fig8:grn}, we selected four representative genes from the hESC dataset under the Specific network and constructed localized GRN subnetworks using their top-13 predicted TF--target pairs ranked by model confidence scores.
For example, scGPL identified the TP53 $\rightarrow$ EPCAM regulatory interaction, which has been experimentally validated by chromatin immunoprecipitation and functional perturbation studies demonstrating direct TP53-mediated regulation of \textit{EPCAM} during cancer progression~\cite{sankpal2009epcam}.
In addition, the predicted TP53 $\rightarrow$ DKK1 and TP53 $\rightarrow$ BMP4 associations are consistent with reported TP53-associated transcriptional programs involved in Wnt signaling modulation, epithelial differentiation, and developmental regulation~\cite{huang2023chipbase}.
scGPL also predicted the STAT3 $\rightarrow$ ANXA3 regulatory link, consistent with experimental evidence showing that STAT3 signaling directly induces \textit{ANXA3} expression and contributes to inflammation and keratinocyte hyperproliferation in psoriasis~\cite{li2025knockdown}.
Similarly, scGPL predicted the TFAP2A regulates ENO1 and BMP4. Recent studies have demonstrated that TFAP2A cooperates with HIF-1$\alpha$ to transcriptionally activate \textit{ENO1} under hypoxic conditions~\cite{Kanoura2025_TFAP2A_HIF_ENO1}, while independent evidence has linked TFAP2A to chromatin-mediated regulation of \textit{BMP4}, contributing to angiogenesis and therapeutic resistance~\cite{Zhang2020_TFAP2A_BMP4_Anlotinib}.
In addition, scGPL predicted the SOX2 $\rightarrow$ PROM1 regulatory link, which is supported by genome-wide CRISPR-Cas9 screening demonstrating that SOX2 directly regulates \textit{PROM1} by binding its promoter and first intron, thereby maintaining CD133 expression and stem cell self-renewal capacity~\cite{savage2025crispr}.
Collectively, these findings indicate that \textit{scGPL} can generalize beyond existing GRN annotations to prioritize biologically meaningful regulatory relationships for downstream experimental validation.
Further case studies and network visualizations are provided in Supplementary Section~D.7 and Figures~S32--S34.

\section{Discussion}
In recent years, GRN inference has evolved beyond conventional GNNs toward more expressive graph representation learning frameworks.
While \textit{GRLGRN} captures local signals from meta-path representations, and \textit{GT-GRN} integrate multimodal gene embeddings, these approaches still primarily train and operate on a single global graph.
\textit{GRNFormer} cleverly addresses this limitation by deconstructing the network into multiple TF-centric subgraphs, though it does not explicitly model the localized context for individual TF-target pairs.
In contrast, \textit{scGPL} adaptively extracts multi-hop neighborhood contexts via dynamic subgraph sampling, uniquely preserving hierarchical structures, pairwise regulatory signals, and holistic topology simultaneously.
Furthermore, \textit{scGPL} flexibly adapts to varying levels of network availability by treating existing interactions as incomplete structural priors, naturally shifting toward expression-based pairwise representations when topological information is severely limited.
This synergistic design prevents the model from merely memorizing specific training topologies, empowering \textit{scGPL} to generalize learned regulatory rules to completely unseen cell types (Supplementary Table S5).
Finally, \textit{scGPL} ensures robust scalability across diverse graph sizes and topological characteristics, as its global message-passing and pair-specific Transformer modules dynamically adapt to varying node counts without relying on fixed global dimensions.

\noindent\textbf{Limitations.}
While \textit{scGPL} demonstrates highly competitive predictive capabilities, it still entails several practical limitations.
The dynamic subgraph sampling strategy introduces additional computational overhead compared with pairwise or graph-based methods, particularly when applied to extremely large candidate interaction spaces.
In addition, its global branch and subgraph modules still rely on the availability of some prior regulatory interactions to provide baseline structural context.
Similar to most existing GRN inference methods, current evaluations are primarily conducted under dataset-specific training and testing settings rather than strict zero-shot cross-dataset generalization.
Therefore, predicting regulatory interactions involving unseen genes or transferring knowledge across distinct biological systems remains an important open challenge.

\section{Conclusion}

We propose \textit{scGPL}, an end-to-end unified framework that synergizes macro-level global topology with fine-grained local interactions via a dynamic subgraph sampling mechanism.
By bridging graph-based and pairwise-based paradigms, \textit{scGPL} leverages data abundance to effectively alleviate sample scarcity and overfitting, thereby consistently outperforming state-of-the-art methods across diverse datasets.
Future work aims to extend this framework by integrating multi-omics modalities (e.g., paired scRNA-seq and scATAC-seq) and leveraging the representational power of large-scale single-cell foundation models. By integrating these multi-modal features through innovative architecture, we anticipate further enhancing the scalability and generalization of our model, ultimately achieving robust zero-shot cross-dataset and cross-species transferability.

\section*{Funding}
This work was supported by the Strategic Priority Research Program of the Chinese Academy of Sciences under Grant No. XDA0460205.

\section*{Data Availability}
{\mdseries scGPL is available on GitHub (\url{https://github.com/talent-hang/scGPL}); the version used in this work is archived on Zenodo (\url{https://doi.org/10.5281/zenodo.21608485}), with evaluation resources for reproducibility.}

\section*{Conflict of Interest}
None declared.

\bibliographystyle{oup-abbrvnat}

\bibliography{reference}

\end{document}